\documentclass[a4paper,11pt]{article}
\usepackage{pos}
\usepackage{hyperref}

\title{The Pulsar Wind Nebulae contribution to gamma-rays}
\ShortTitle{The PWNe contribution to gamma-rays}

\author*[a,b]{Barbara Olmi}

\affiliation[a]{INAF, Osservatorio Astronomico di Palermo\\
  Piazza del Parlamento 1, Palermo, Italy}

\affiliation[b]{INAF, Osservatorio Astrofisico di Arcetri,\\
Largo Enrico Fermi 5, Firenze, Italy}

\emailAdd{barbara.olmi@inaf.it}

\abstract{
Pulsar Wind Nebulae (PWNe) shine at multi-wavelengths and are expected to constitute the largest class of gamma-ray sources in our Galaxy. They are known to be very efficient particle accelerators: the Crab nebula, the PWNe class prototype, is the unique firmly identified leptonic PeVatron of the Galaxy to date, and most of the PeVatrons recently detected by LHAASO appear to be compatible with a pulsar origin.
PWNe have been proved to be associated with the formation of misaligned X-ray tails and TeV halos, as sign of an efficient escape of energetic particles from the PWN into the surrounding medium.

With the advent of the Cherenkov Telescope Array we expect that $\sim 200$ new PWNe will be detected.
Being able to correctly model their multi-wavelength spectral properties, spatial and spectral morphology at gamma-rays is then topical today. 
This in particular means we should be able to account for their different evolutionary phases, and to correctly determine the influence they have on the spectral properties of the source. This indeed reflects directly on the expectation of how many PWNe will be detected at gamma-rays.

Finally, the identification of PWNe in future gamma-ray data, not only is relevant for their scientific importance, but also to allow for the identification of less prominent sources that might be hidden by the background of non-identified PWNe.

}

\FullConference{%
  7th Heidelberg International Symposium on High-Energy Gamma-Ray Astronomy (Gamma2022)\\
    4-8 July 2022\\
  Barcelona, Spain
}

\begin{document}
\maketitle

\section{Introduction}
A Pulsar Wind Nebula (PWN) is produced by the interaction of the relativistic, cold and magnetized outflow emanated by the pulsar in its spin-down process with the ambient medium.
This outflow, named pulsar wind, is made by magnetic field and particles, mostly (or totally) leptons plus a (possible) hadronic component \citep{Amato:2003}. 
The composition of the wind is still  a matter of debate and ongoing investigation, especially in the case of the Crab nebula \citep{Amato_Olmi:2021}, where new spectral information have been recently added at extremely high energies ($E>100$ TeV, EHE) thanks to LHAASO \citep{LHAASO_crab:2021}. 
This is where the hadronic contribution is expected to  emerge from the leptonic one, and eventually becomes detectable \cite{Amato_Olmi:2021}.

For young enough systems the ambient medium is likely the remnant of the supernova explosion, the un-shocked ejecta.
The interaction of the relativistic pulsar outflow with the slowly expanding supernova remnant (SNR) ejecta induces the formation of strong magneto-hydrodynamic (MHD) shock, that literally terminates the pulsar wind (called termination shock, TS). 
At the crossing of the TS the wind plasma is heated and slowed down, while magnetic field is partially dissipated.
The TS is also the location at which we believe particle acceleration takes place.

A different scenario describes old enough systems. Considering that the pulsar population is characterized by an average kick velocity of order of $100-500$ km s$^{-1}$ (e.g. \cite{FGK:2006,Verbunt:2017}), much larger than the typical expansion velocity of the SNR shell, many pulsars will escape their SNR bubble at a certain point. Then they will directly interact with the interstellar medium (ISM), where the pulsar velocity generally exceeds, by a factor of $\gtrsim 10$, the local sound speed; as a consequence a bow shock forms at the PWN boundary with the ISM, mediating the supersonic motion of the pulsar in the medium. The PWN is then reshaped in a cometary-like fashion, and it is known as a bow shock nebula.

Young systems are detected at an extremely broad range of energies, from radio to gamma-rays, with a fully non-thermal spectrum where the main emitting component is produced via synchrotron radiation.
Higher energy emission is instead produced by inverse Compton scattering (ICS) between relativistic particles of the wind and different target photons (with mayor contributions from the cosmic microwave background and infrared photons).
The appearance of a PWN at multi-wavelengths may change consistently as time passes by, with old systems likely characterized by the predominance of ICS over synchrotron, both due to the lower energy injection of the pulsar at that stage, and to the synchrotron cooling.
We can easily compute the synchrotron lifetime of a particle emitting a 1 keV photon in a 50$\,\mu$G nebular magnetic field as:
\begin{equation}\label{tausync}
    \tau_{\rm{sync}}\simeq 156 \, \left( \frac{B_{\mu\mathrm{G}}}{50} \right)^{-3/2} \left( \frac{ E_{\rm{ph, keV}}}{1} \right)^{-1/2}\rm{yr}\,,
\end{equation}
clearly showing that a PWN is visible at X-rays approximately only for the period in which the pulsar is enough energetic to inject particles able to produce X-ray photons; the life time of X-ray emitting particles in the nebula field is indeed very limited, and as such X-rays only trace the freshly injected plasma.
On the contrary, particles responsible for the low energy emission, have much longer lifetimes against synchrotron losses. The lifetime of a particle producing a $10^{11}\,$Hz (radio) photon in the same magnetic field as before, is in fact three order of magnitude larger: $\tau_{\rm{sync}}\simeq 2\times 10^5\,$yr. 
A PWN will then remain visible in the radio band for much longer times than at X-rays.

The persistence of the emission at very high energies is a different story.
ICS needs much less energetic electrons than synchrotron to produce high energy photons. 
For example, a 10 TeV electron (or positron) is sufficient to generate a 1 TeV photon via ICS with a CMB photon as a target. 
On the contrary, the same 10 TeV lepton spiralling in a 50  $\mu$G field, only produces a $\sim 0.1$~keV photon via synchrotron radiation.
The long-living radio electrons in the PWN then become the main actors to produce gamma-ray emission, especially in old systems, where higher energy particles have disappeared.
The consequence is that the PWN will also shine at gamma-rays for a very long time.
If we assume $\sim100$ kyr as an average estimate of the lifetime at gamma-rays of a PWN, and a rate of birth for Galactic pulsars of 1/100 yr \citep{FGK:2006}, the number of PWNe expected at gamma-rays is then of $\sim1000$.
Of them, a large part are likely to be detected only at gamma-rays, being too old to still emit at X-rays, while the diffuse radio emission of an old object might be difficult to be revealed due to the lack of sensitivity to the required angular scales of most of the radio interferometers.

\section{The different phases of PWNe evolution}
\begin{figure}
\centering
	\includegraphics[width=.99\textwidth]{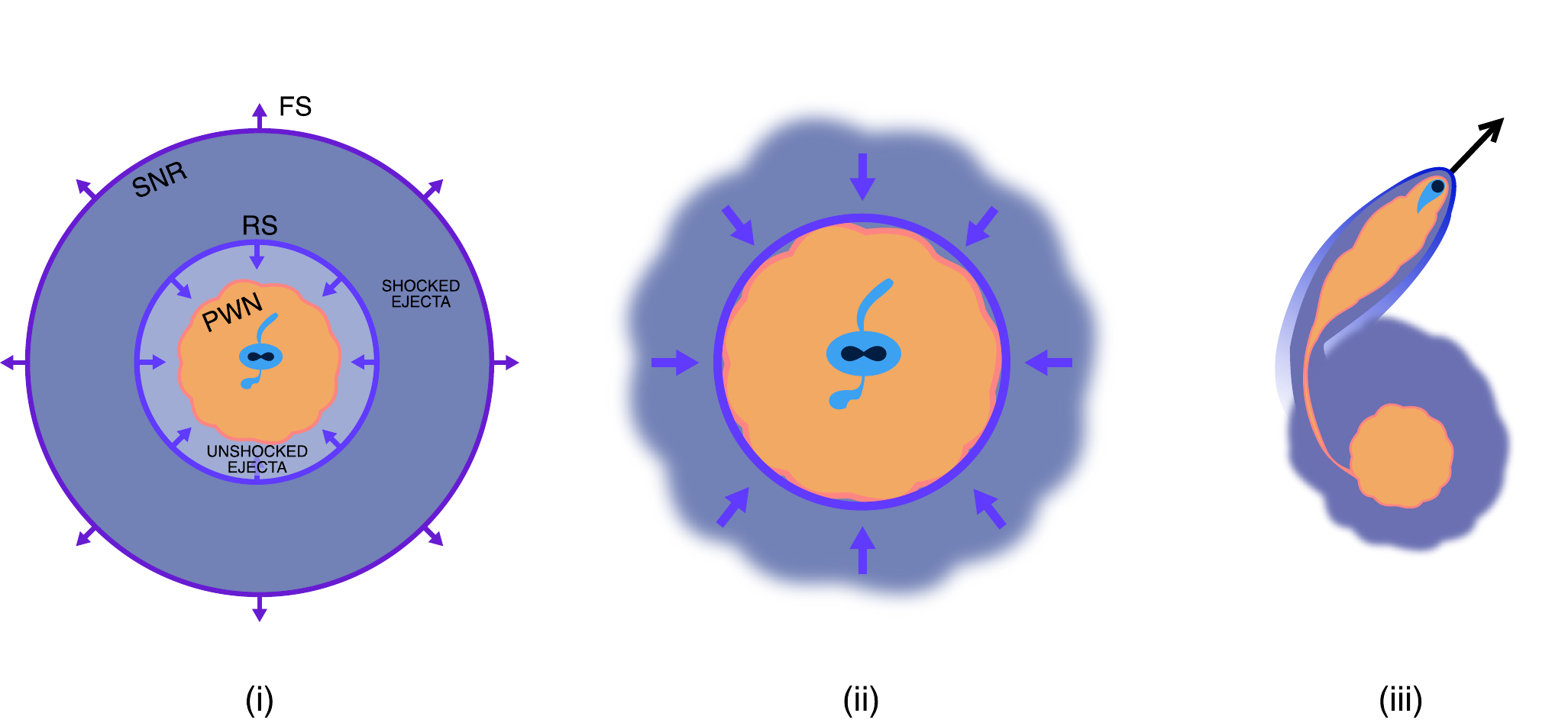}
    \caption{Sketch showing the three main evolutionary stages of PWNe. (i) Free expansion phase: the PWN expands with mild acceleration in the free expanding SNR (the boundary is the forward shock FS). The interaction between the PWN and the SNR has not yet established, the supernova reverse shock (RS) is still travelling towards the center of explosion, and only a part of the ejecta has been shocked. (ii) Reverberation phase: the RS hits the PWN boundary on its way and interaction stars. The outcome of this interaction depends on the energetics of the whole system. (iii) Late phase: the PWN leaves the SNR and becomes a bow shock nebula. A relic bubble of radio particles may be left behind the runaway pulsar. Figure adapted from \cite{Olmi_Bucciantini:2022}.} 
    \label{fig:PWNevo}
\end{figure}
%
The evolution of a PWN can be roughly divided into three main phases, shown in the sketches of Fig.~\ref{fig:PWNevo}: (i) the free expansion phase; (ii) the reverberation phase; (iii) the late or bow shock phase. For a recent review about the different PWNe evolutionary phases, available models, and comparison with observation, we refer the reader to \cite{Olmi_Bucciantini:2022}.

In the first phase the PWN expands, with mild acceleration, in the freely expanding SNR, and a direct interaction between the two has not been established yet \citep{Reynolds_Chevalier:1984}.
This phase was largely investigated in the past, using a number of different models and approaches.
Impressive results came from relativistic MHD simulations, especially thanks to  recent 3D models \citep{Porth:2014, Olmi:2016}. 
As one of their main results it is worth to mention the possibility to reach a magnetization ($\sigma$) larger than unity in the pulsar wind. This is an important parameter, setting the initial distribution of the wind energy flux into magnetic field ($B$) and particles, given by:  $\sigma=B^2/[4\pi\Gamma_w \rho c^2]$ (with $\Gamma_w$ the wind  Lorentz factor and $\rho$ the plasma density). 
For the first time, 3D models reduced consistently the distance between theoretical expectations from pulsar theories (requiring $\sigma \gg 1)$ and PWNe models (the famous \textit{sigma paradox}), thanks to the development of efficient magnetic dissipation in the nebula (especially conveyed by the kink instability).
%
\begin{figure}
\centering
	\includegraphics[width=.69\textwidth]{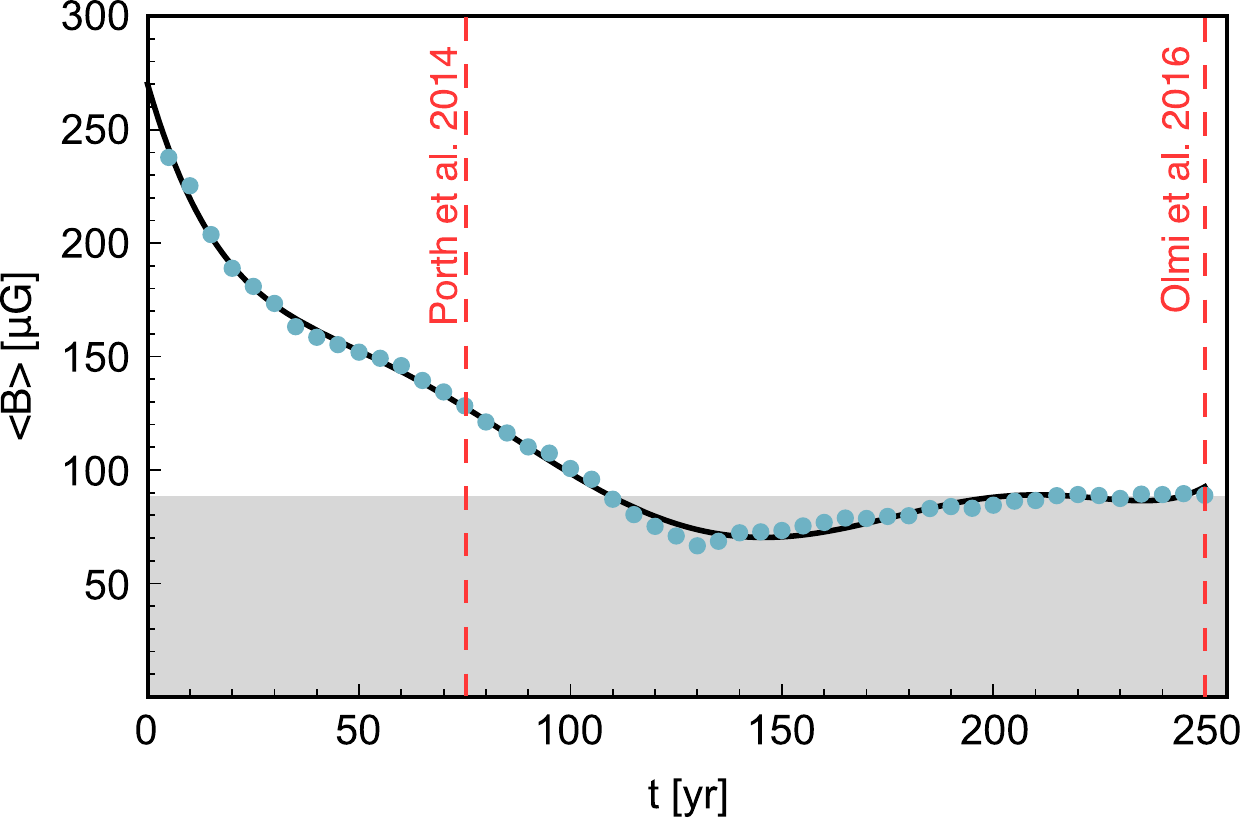}
    \caption{Time evolution of the average magnetic field in the 3D simulation of the Crab nebula by \cite{Olmi:2016}. The red-dashed vertical lines indicate the final time of that simulation compared to the longest run performed by \cite{Porth:2014}, where the self-similar evolution was not yet reached and the magnetic field is still visibly decreasing.} 
    \label{fig:Bfield3D}
\end{figure}
%
Nevertheless the magnetic field from 3D simulations is still on average too low to correctly reproduce the high energy and ICS spectral properties of the nebula; in \cite{Olmi:2016} a value of $90\,\mu$G was found for the Crab nebula when the self-similar expansion has been reached, to be compared with the expected $\sim 112-150\,\mu$G (see Fig.~\ref{fig:Bfield3D}). 
This might indicate that at injection the pulsar wind must have a  magnetization well beyond unity.

The main limitation of MHD models is the lack of a consistent radiative treatment. Particles are hidden in the thermal plasma and their spectral properties are only computed with a post-processing recipe at a chosen point of the evolution.
Radiative modelling is instead possible using one-zone models. The PWN is treated as a uniform bubble subject to energy losses (adiabatic + radiative) and possibly particles leakage, interacting with the surrounding SNR. 
Their main assumption is the thin-shell approximation, for which the PWN radius can be confused with that of the thin shell of swept up ejecta that accumulates at the PWN boundary during its initial expansion \cite{Jun:1998,Gelfand:2009}.
The time evolution of the shell radius is coupled to the description of the particles distribution function, considering both injection and losses, and from that one derives the PWN multi-wavelength spectrum.
Despite their simplified representation of the PWN, one-zone models have been shown to provide a good description of the macroscopic properties of young systems; moreover they are not much demanding in terms of numerical resources, thus they are widely used, especially  for population studies that require long evolution of a large number of sources (see e.g. \citep{Torres:2014, Fiori:2022}).

The main problem with one-zone models becomes apparent when trying to model  the PWN evolution beyond the free-expansion phase.
When the SNR reverse shock reaches the PWN boundary, the interaction with the SNR starts. This phase is known as reverberation, since the PWN may experience a number of compressions and re-expansions.
Recently \cite{Bandiera:2022} shows that one-zone models generally over-predict the compression of the nebula in this phase, possibly leading to the burn-off of large part of the emitting particles due to the enhanced magnetic field, changing substantially the spectral properties of the source. 
The main cause of this is to be found in the pressure profile assumed in the SNR: this is generally shaped using the central pressure of the Sedov solution or the pressure at the SNR forward shock, arbitrarily scaled. In both cases the pressure turns out to be largely overestimated, then exerting much more pressure on the PWN than expected, that translates in general into a stronger compression.
Solving this issue is particularly relevant for population studies, since the modification of the spectral properties of the population might introduce important biases in the number of expected sources in given observational band. 
A first attempt to extend the reliability of one-zone models beyond reverberation has been made in \cite{Fiori:2022}, in the context of the study of the gamma-ray properties of the Galactic PWNe population.
A more refined model has been recently presented in \cite{Bandiera:2022}, though limited to the non-radiative case.

After reverberation PWNe are likely characterized by strong-asymmetries, both as direct consequence of the compression, of border instabilities (as Rayleigh-Taylor ones), of the interaction with density gradients in the ambient medium, as well as as due to the pulsar proper motion. 
To date we only have identified and characterized a bunch of those systems, thus it is very difficult to develop models with geenral valence for this complex phase. A famous example is G327.1-1.1 \cite{Temim:2015}.

Instead we have detected a greater number of PWNe associated with runaway pulsars, the bow shock nebulae.  These systems are likely associated with faint pulsars (not all, as the case of the powerful Lighthouse nebula \cite{Pavan:2014}), and to date no one has been seen directly at gamma-rays.
On the other hand they have been shown to be sources of high energy particles escaping from the bow shock head, observed in the form or elongated, thin and misaligned X-ray tails, extending in the direction opposite to the direction of motion of the pulsar \cite[see][and references therein.]{Olmi_Bucciantini:2019c}, or extended TeV halos \cite{Abeysekara:2017}.

\section{Contribution to gamma-rays}
PWNe are expected to contribute to Galactic gamma-rays in both their first two evolutionary phases, while, as already mentioned, no gamma-ray emission has been detected so far from bow shock PWNe.
This might be due to the limited spatial extension of the bow shock head (tipically $\sim$ few$\,\times 10^{16}$ cm) as well as to their low residual luminosity.
This means that bow shock nebulae should be statistically irrelevant at gamma-rays. On the other hand their are associated to two important additional sources of gamma-rays: relic nebulae, the radio bubbles left behind a pulsar escaping its own PWN bubble; and TeV halos. 

To date only a limited number of PWNe have been firmly identified in gamma-ray surveys. Out of the 24 extended sources detected in the H.~E.~S.~S. Galactic Plane Survey \citep{HESSGPS:2018}, only 14 are those firmly identified as PWNe, thanks to the presence of a multi-wavelength counterpart, while 10 remain not associated.
Similarly, in the Fermi-LAT 3FGL catalog \citep{3FGL:2015}, unidentified sources are $\sim 20$\% of the total.
Recalling the estimate of the number of expected Galactic PWN  at gamma-rays ($\sim 1000$), it seems reasonable that many -- if not all -- are PWNe, meaning that these sources can alone represent around 40\% of the total gamma-ray sources in the Galaxy.
%
\begin{figure}
\centering
	\includegraphics[width=.5\textwidth]{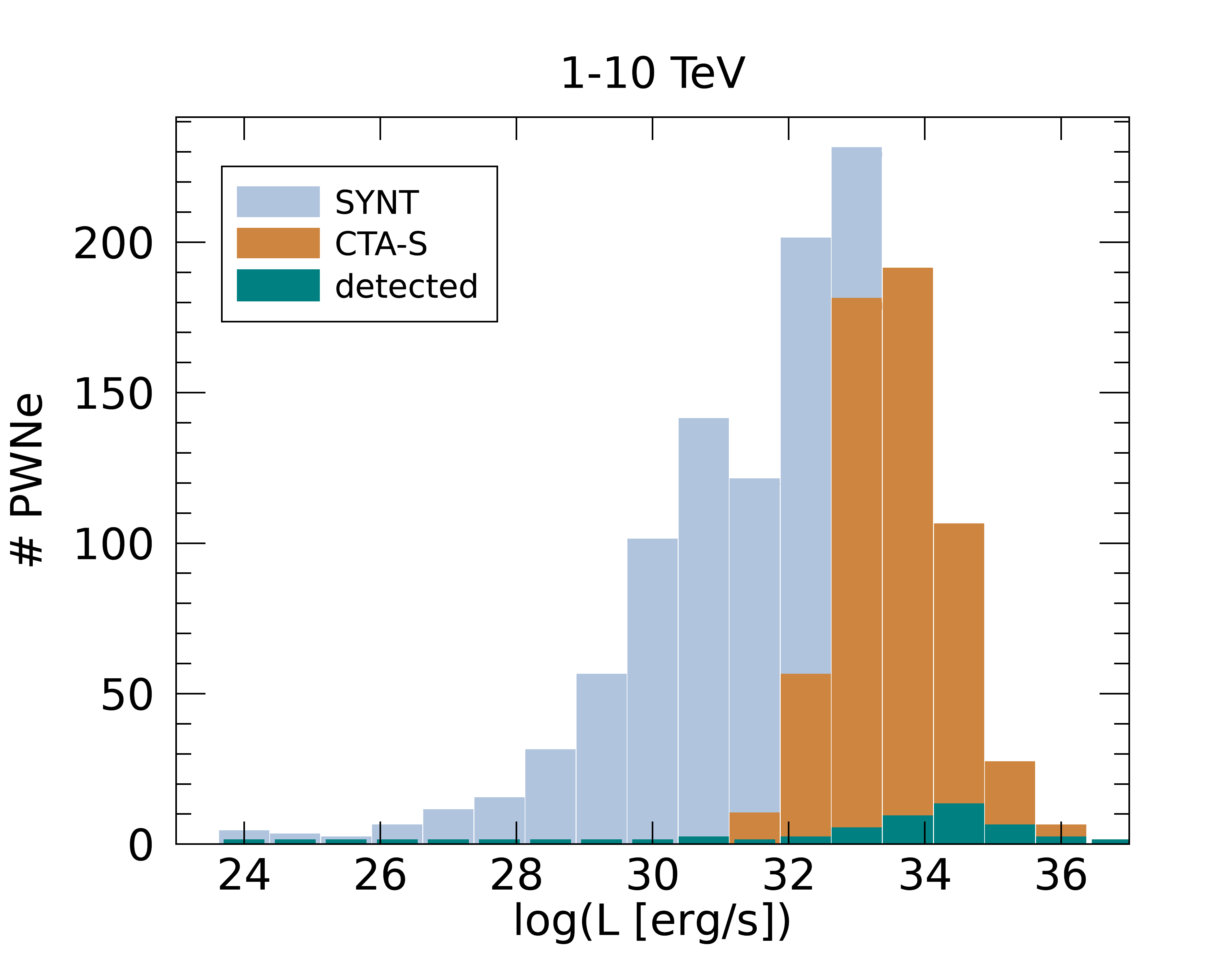}
 \hspace{-0.7cm}
 	\includegraphics[width=.5\textwidth]{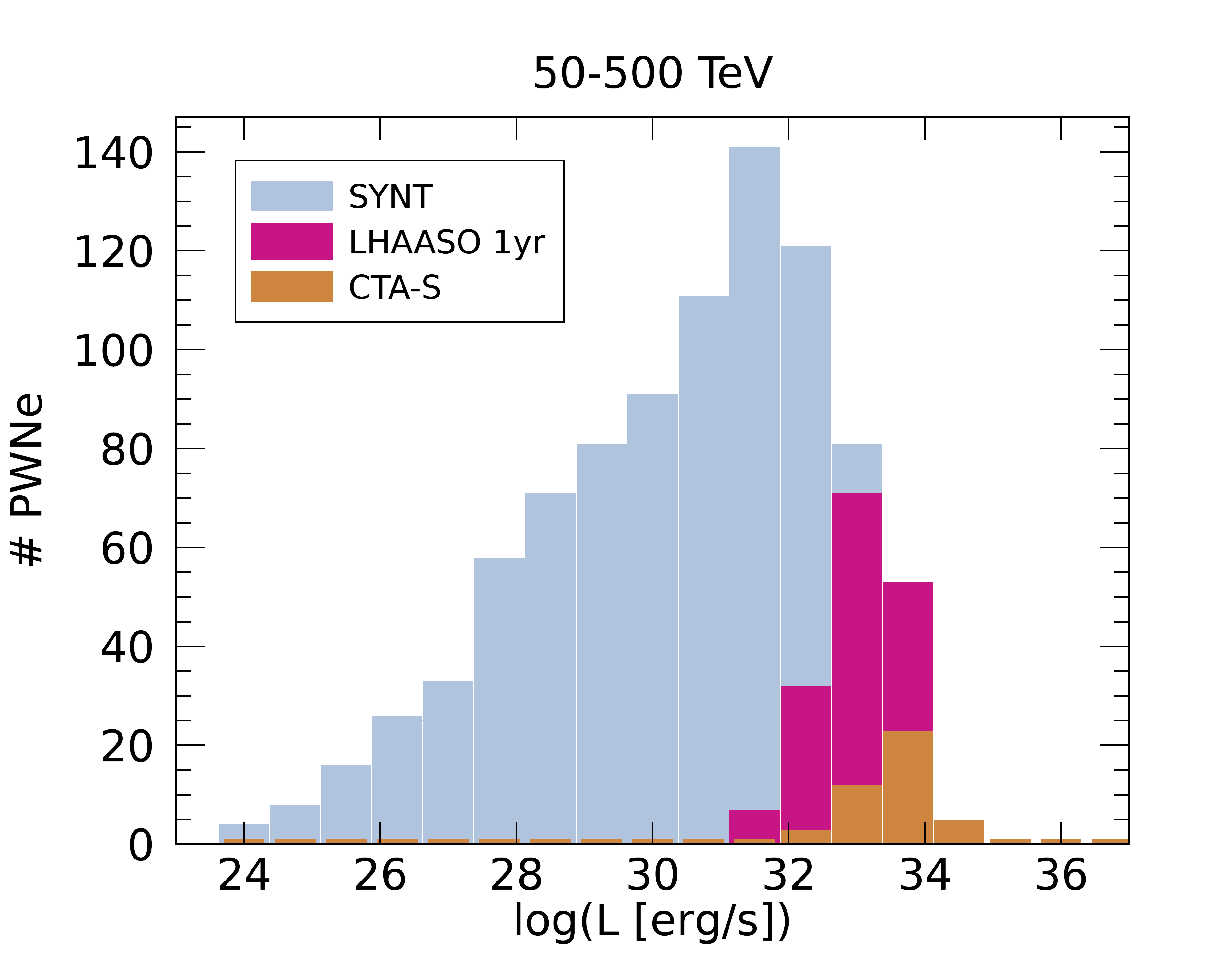}
    \caption{Histogram of the theoretical expected number of PWNe in the Galaxy, per luminosity beam (in light blue) and in two different energy bands (1-10 TeV left panel; 50-500 TeV right panel). The synthetic population is the one presented in \cite{Fiori:2022}, obtained with a one-zone model based on the association of the populations of Galactic gamma-ray pulsars and core-collapse SNRs.
    In light brown it is shown the number of expected detections with CTA-South, considering a threshold in flux of $0.5\times10^{-13}$ erg$\,$s$^{-1}\,$cm$^{-2}$ \citep{Remy:2022}. The present number of detected sources at 1-10 TeV is shown in green in the left panel, while in the right panel the magenta bars indicate the sources above the LHAASO detection threshold with 1 yr integration (see \cite{Fiori:2022} for details).} 
    \label{fig:PWNisto}
\end{figure}
%
We now believe that PWNe may also dominate the EHE domain, since 11 out of the 12 PeV sources recently detected by LHAASO \cite{LHAASO_12s:2021} are compatible with being powered by a pulsar \cite{de-Ona-Wilhelmi:2022}.

With the advent of new gamma-ray facilities, especially the high resolution and high sensitivity new generation of IACTs\footnote{Imaging Atmosferic Cherenkov Telescopes.} (as the ASTRI Mini-Array and the Cherenkov Telescope Array, CTA), we will have access to a huge amount of new data. 
In Fig.~\ref{fig:PWNisto} we can see that the number of theoretically detectable PWNe in the 1-10TeV energy band with CTA will increase from the $\sim30$ known at present to $\sim 550$. 
Considering a conservative fraction of  1/3 of systems that will be actually identified on top of the theoretical detectable, this means that we expect around $\sim 200$ new PWNe to be revealed in the first Galactic Plane Survey by CTA.

\section{Summary and Perspectives}
Being the most numerous expected sources in the upcoming gamma-ray data, modelling PWNe becomes topical today.
Due to the lack of high-resolution data and morphological information at gamma-rays, so far models have been mostly developed to match lower energies, where the comparison is most significant.
To reproduce the PWNe contribution to gamma-rays, models must deal with the various evolutionary stages, since most of the sources will be old. As discussed earlier, this means than more attention needs to be paid than in the past especially to the reverberation phase, since its influence on the spectral history of a source cannot be ignored, being the gamma-ray emission significantly dependent on the past injection properties.
This is particularly important for population studies, since a wrong determination of the spectral properties of PWNe at gamma-rays might change significantly the number of expected sources in the upcoming gamma-ray data.

Moreover the identification of evolved pulsars (and pulsar TeV halos as well) is not only important for their scientific relevance, but also since they need to be taken into account carefully as major contributors to the Galactic gamma-ray background in the detection of other sources, as the hadronic PeVatrons, primary goal of all the upcoming facilities.

Thanks to the upcoming generation of IACTs we also expect to finally address the possible presence of hadrons in the pulsar wind, through precise measurements of the VHE-EHE spectrum, where the presence of hadrons is expected to emerge from the dominant leptonic component.

\acknowledgments
The author wish to acknowledge all the colleagues that contribute to the work presented in this proceeding, in particular: Elena Amato, Rino Bandiera,  Niccolò Bucciantini, Luca Del Zanna, Michele Fiori, Andrea Mignone, Diego F. Torres.
The author also acknowledges support by INAF grants MAINSTREAM 2018, SKA-CTA, PRIN-INAF 2019 and by ASI-INAF n.2017-14-H.O.

\bibliographystyle{JHEP}
\bibliography{biblio}

\end{document}